\documentclass[aps,preprint]{revtex4}%
\usepackage{amsfonts}
\usepackage{amsmath}
\usepackage{amssymb}
\usepackage{graphicx}%
\begin{document}
\title[ ]{The Blackbody Radiation Spectrum Follows from Zero-Point Radiation and the
Structure of Relativistic Spacetime in Classical Physics}
\author{Timothy H. Boyer}
\affiliation{Department of Physics, City College of the City University of New York, New
York, New York 10031}
\keywords{Blackbody radiation; thermal equilibrium; scaling symmetry; classical
electromagnetism, Rindler frame, relativistic spacetime, conformal symmetry}
\pacs{}

\begin{abstract}
The analysis of this article is entirely within classical physics. \ Any
attempt to describe nature within classical physics requires the presence of
Lorentz-invariant classical electromagnetic zero-point radiation so as to
account for the Casimir forces between parallel conducting plates at low
temperatures. \ Furthermore, conformal symmetry carries solutions of Maxwell's
equations into solutions. \ In an inertial frame, conformal symmetry leaves
zero-point radiation invariant and does not connect it to
non-zero-temperature; time-dilating conformal transformations carry the
Lorentz-invariant zero-point radiation spectrum into zero-point radiation and
carry the thermal radiation spectrum at non-zero temperature into thermal
radiation at a different non-zero-temperature. \ However, in a non-inertial
frame, a time-dilating conformal transformation carries classical zero-point
radiation into thermal radiation at a finite non-zero-temperature. \ By taking
the no-acceleration limit, one can obtain the Planck radiation spectrum for
blackbody radiation in an inertial frame from the thermal radiation spectrum
in an accelerating frame. \ Here this connection between zero-point radiation
and thermal radiation is illustrated for a scalar radiation field in a Rindler
frame undergoing relativistic uniform proper acceleration through flat
spacetime in two spacetime dimensions. \ The analysis indicates that the
Planck radiation spectrum for thermal radiation follows from zero-point
radiation and the structure of relativistic spacetime in classical physics.

\end{abstract}
\maketitle

\section{Introduction}

Although the universal spectrum of blackbody radiation has historically been
connected with statistical mechanics and quantum theory, the present article
makes the connection with zero-point radiation and the structure of
relativistic spacetime. \ It is suggested that \ classical thermal radiation
in a general static coordinate system is the one-parameter stationary spectrum
of isotropic random classical radiation obtained from the scale-invariant
spectrum of zero-point radiation by the time-dilating conformal transformation
which preserves the wave equation and carries radiation normal modes into
normal modes.\cite{PR2010}\cite{AJP2011} \ Although it is suggested that this
definition of thermal radiation can be extended to a general spacetime, it is
illustrated here only for a scalar radiation field in flat spacetime in two
spacetime dimensions. \ It is pointed out that an inertial frame is such a
specialized coordinate frame that a time-dilating conformal transformation is
identical with a space-dilating conformal transformation, and that zero-point
radiation in an inertial frame remains invariant under all conformal
transformations and so does not allow us to obtain the blackbody spectrum at
non-zero temperature from the Lorentz-invariant zero-point radiation spectrum.
\ However, the thermal radiation spectrum in an inertial frame can be obtained
by going to a relativistic accelerating frame (a Rindler frame), applying a
time-dilating conformal transformation to zero-point radiation in that
coordinate frame, and then carrying the spectrum back to an inertial frame by
taking the no-acceleration limit. \ Work related to the present article has
been published in the Physical Review D\cite{PR2010} and in the American
Journal of Physics\cite{AJP2011}; however, in these articles, the connection
with the conformal group is absent. \ 

The work described here is completely at variance with the suggestions of the
physicists in the early 20th century that classical physics leads inevitably
to the Rayleigh-Jeans spectrum as the spectrum of blackbody
radiation.\cite{Eisberg} \ The physicists of that period were aware neither of
classical zero-point radiation nor of the importance of relativity and so used
nonrelativistic statistical mechanics or scattering of radiation by
nonrelativistic nonlinear systems when determining the spectrum of thermal
equilibrium. \ In the present article, we are replacing these nonrelativistic
calculations by symmetry considerations in relativistic spacetime. \ In line
with the suggestions here regarding the importance of relativistic physics for
radiation equilibrium, it has been shown that the zero-point radiation
spectrum is invariant under scattering by the relativistic scattering system
of a relativistic charged particle in a Coulomb potential.\cite{FOP2010}
\ This result is again in complete contrast with the scattering results
involving nonrelativistic nonlinear mechanical systems.\cite{VanV}

\section{Preliminaries}

\subsection{Conformal Symmetry and Fundamental Constants}

Electromagnetic theory is invariant not only under Lorentz transformations but
also under the still larger group of conformal transformations.\cite{BC1909}
\ The conformal transformations in an inertial frame include both the
scale-changing dilatations (which will be used in the present article), as
well as the special conformal transformations which correspond to local
changes of the spacetime scale.\cite{Kastrup} \ Sometimes the dilatations of
the conformal group have been referred to as $\sigma_{ltU^{-1}}$-scale
changes\cite{SCALING} to emphasize that these changes transform length and
time with the same multiplicative constant while transforming the energy with
the inverse of the multiplicative constant. \ Thus in an inertial frame, if
under a dilatation the length $l$ is mapped to $\overline{l}=\sigma l$ where
$\sigma$ is a positive real number, then the time is mapped as $t\rightarrow
\overline{t}=\sigma t,$ and the energy $U$ mapped as $U\rightarrow\overline
{U}=U/\sigma.$ \ Thus the dilatations of the conformal group preserve the
values of the fundamental constant $c$ (the speed of light in vacuum,
involving length divided by time), the constant $a_{S}/k_{B}^{4}$ (Stefan's
constant of blackbody radiation divided by Boltzmann's constant to the fourth
power, involving the inverse third power of energy times length), and the
constant $e$\ (the charge of the electron, involving the square-root of energy
times length).

\subsection{Metrics}

In this article, we will discuss not electromagnetic radiation but rather the
analogue system of relativistic scalar radiation in two spacetime dimensions
in flat spacetime. \ In an inertial frame, the coordinates $x^{0}=ct,$
$x^{1}=x$ can be chosen as geodesic coordinates and the spacetime metric
$ds^{2}=g_{\mu\nu}dx^{\mu}dx^{\nu}$ takes the Minkowski form%
\begin{equation}
ds^{2}=dx^{0}dx^{0}-dx^{1}dx^{1}=c^{2}dt^{2}-dx^{2} \label{eM2}%
\end{equation}
Thus in an inertial frame, the time coordinate $t$ is connected to the
interval length by the universal constant $c$. \ 

A Rindler coordinate frame\cite{Rindler} in flat spacetime involves
time-constant proper acceleration at each point of the static coordinate
frame. We can pass from an inertial frame to a Rindler frame by relating the
time and space coordinates $ct,x$ in an inertial frame to $\eta,\xi$ in the
Rindler frame as%
\begin{equation}
ct=x^{0}=\xi\sinh\eta\label{eR1}%
\end{equation}%
\begin{equation}
x=x^{1}=\xi\cosh\eta\label{eR2}%
\end{equation}
The metric is then given by%
\begin{equation}
ds^{2}=c^{2}dt^{2}-dx^{2}=\xi^{2}d\eta^{2}-d\xi^{2} \label{eMR}%
\end{equation}
In this metric involving $\eta$ and $\xi$, the constant $c$ no longer appears
explicitly and the time parameter $\eta$ does not carry any units. \ 

\subsection{Scalar Radiation Field}

In an inertial frame, the Lagrangian for a massless relativistic real scalar
field $\phi$ is given by
\begin{equation}
\mathcal{L=}\frac{1}{8\pi}\partial_{\mu}\phi\partial^{\mu}\phi=\frac{1}{8\pi
}\left[  \left(  \frac{\partial\phi}{\partial ct}\right)  ^{2}-\left(
\frac{\partial\phi}{\partial x}\right)  ^{2}\right]  \label{eL}%
\end{equation}
From the Lagrangian (\ref{eL}), it follows that the wave equation
$\partial_{\mu}[\partial\mathcal{L}/\partial(\partial_{\mu}\phi)]-\partial
\mathcal{L}/\partial\phi=0$ for the field is%
\begin{equation}
0=\frac{\partial^{2}\phi}{\partial(ct)^{2}}-\frac{\partial^{2}\phi}{\partial
x^{2}} \label{W1}%
\end{equation}
while the stress-energy-momentum tensor is $\mathcal{T}^{\mu\nu}%
=[\partial\mathcal{L}/\partial(\partial_{\mu}\phi)]\partial^{\nu}\phi
-g^{\mu\nu}\mathcal{L}$ so that\ the energy density is%
\begin{equation}
u=\mathcal{T}^{00}=\frac{1}{8\pi}\left[  \frac{1}{c^{2}}\left(  \frac
{\partial\phi}{\partial t}\right)  ^{2}+\left(  \frac{\partial\phi}{\partial
x}\right)  ^{2}\right]  \label{eED1}%
\end{equation}
and the momentum density is
\begin{equation}
\mathcal{T}^{01}=-\frac{1}{4\pi}\frac{\partial\phi}{\partial ct}\frac
{\partial\phi}{\partial x} \label{emd}%
\end{equation}

In $n$ spacetime dimensions, the energy density corresponding to Eq.
(\ref{eED1}) would have the dimensions of $energy/(length^{n-1})$. Therefore
the square of the field $\phi^{2}$ has dimensions of $energy/(length^{n-3}).$
\ We will consider a scalar field in two spacetime dimensions $n=2$ where
energy can be measured in units of inverse length and the square of the field
$\phi^{2}$ involves energy times length and so can be treated as dimensionless.

By introducing the coordinate transformations of Eqs. (\ref{eR1}) and
(\ref{eR2}) into the wave equation (\ref{W1}) together with the assumption of
scalar behavior $\phi(ct,x)=\varphi(\eta,\xi)$ for the relativistic radiation
field in two spacetime dimensions, we can obtain the wave equation for the
radiation field $\varphi(\eta,\xi)$ in the Rindler frame as\cite{AJP2011}%
\begin{equation}
0=\frac{1}{\xi^{2}}\frac{\partial^{2}\varphi}{\partial\eta^{2}}-\frac
{\partial^{2}\varphi}{\partial\xi^{2}}-\frac{1}{\xi}\frac{\partial\varphi
}{\partial\xi} \label{eWR1}%
\end{equation}

\subsection{Normal Modes}

The normal modes of the scalar field $\phi_{k}$ are solutions of the wave
equation with harmonic time dependence%
\begin{equation}
\phi_{k}(ct,x)=\operatorname{Re}\phi_{(k)}(x)\exp[-i|k|ct] \label{eNM1}%
\end{equation}
In an inertial frame, the normal modes of Eq. (\ref{W1}) are solutions of the
scalar Helmholtz equation%
\begin{equation}
\frac{\partial^{2}\phi_{(k)}}{\partial x^{2}}+k^{2}\phi_{(k)}=0 \label{eH1}%
\end{equation}
which can be chosen as $\phi_{(k)}(x)=\exp[ikx],$ so that the general normal
mode involves running wave solutions%
\begin{equation}
\phi(ct,x)=\operatorname{Re}\exp[ikx-i|k|ct] \label{M1}%
\end{equation}
where $k$ can be positive or negative.$.$

In the Rindler frame, the normal modes $\varphi_{\kappa}$ of the wave equation
will not take the simple running wave form of Eq. (\ref{M1}). The spatial
dependence is more complicated and must satisfy%
\begin{equation}
0=+\frac{\partial^{2}\varphi_{(\kappa)}}{\partial\xi^{2}}+\frac{1}{\xi}%
\frac{\partial\varphi_{(\kappa)}}{\partial\xi}+\frac{1}{\xi^{2}}\kappa
^{2}\varphi_{(\kappa)} \label{eH2}%
\end{equation}
with solution $\varphi_{(\kappa)}(\xi)=\xi^{\pm i\kappa}=\exp[\pm i\kappa
\ln\xi]$ , so that in the Rindler frame the general normal mode in two
spacetime dimensions is%
\begin{equation}
\varphi_{\kappa}(\eta,\xi)=\operatorname{Re}\exp[i\kappa\ln\xi-i|\kappa|\eta]
\label{NM2}%
\end{equation}
where the parameter $\kappa$ can be positive or negative corresponding to
waves moving in either the positive or negative direction. \ 

\subsection{Time-Dilating Conformal Transformations}

When discussing thermal radiation, we will be interested in time-dilating
conformal transformations which carry radiation normal modes into normal
modes. \ Now conformal transformations of the metric\cite{Rohrlich} involve
those mappings $x^{\mu}\rightarrow\overline{x}^{\mu}=h^{\mu}(x^{\nu})$ which
transform the metric in the form%
\begin{equation}
g_{\mu\nu}(x^{\alpha})=\Omega^{2}(\overline{x}^{\beta})g_{\mu\nu}(\overline
{x}^{\beta}) \label{eC1}%
\end{equation}
These transformations leave the wave equation invariant for a tensor field.
\ However, we are not interested in conformal transformations of the metric or
a change in coordinate system. \ Rather we will want to maintain the
coordinate system and the metric while making an active transformation to a
new radiation field in the original coordinate system. \ 

A time-dilating conformal transformation takes the time coordinate into a
multiple $\sigma$ of itself, and requires that the transformation of the
spatial coordinates are such that the wave equation remains invariant and the
metric transforms as in Eq. (\ref{eC1}). In an inertial frame in two spacetime
dimensions, a time-dilating conformal transformation is simply a uniform
dilatation of both the time and space coordinates, $t\rightarrow\overline
{t}=\sigma t$, $x\rightarrow\overline{x}=\sigma x$, while the radiation field
$\phi$ transforms as a scalar field under the coordinate transformation
$\overline{\phi}(c\overline{t},\overline{x})=\phi(ct,x)$ . This transformation
carries the wave equation into itself%
\begin{equation}
\frac{1}{c^{2}}\frac{\partial^{2}\overline{\phi}}{\partial\overline{t}^{2}%
}-\frac{\partial^{2}\overline{\phi}}{\partial\overline{x}^{2}}=\frac{1}%
{\sigma^{2}}\left(  \frac{1}{c^{2}}\frac{\partial^{2}\phi}{\partial t^{2}%
}-\frac{\partial^{2}\phi}{\partial x^{2}}\right)  =0 \label{eT1}%
\end{equation}
and carries plane waves of frequency $c|k|$ into plane waves of frequency
$c|k|/\sigma$ as%
\begin{align}
\overline{\phi}(c\overline{t},\overline{x})  &  =\phi(ct,x)\nonumber\\
&  =\phi(c\overline{t}/\sigma,\overline{x}/\sigma)\nonumber\\
&  =\operatorname{Re}\exp\left[  i\frac{k}{\sigma}\overline{x}-i|\frac
{k}{\sigma}|c\overline{t}\right]  \label{F1}%
\end{align}
This uniform dilation corresponds to a conformal transformation in an inertial
frame since%
\begin{equation}
ds^{2}=c^{2}dt^{2}-dx^{2}=\sigma^{-2}[c^{2}d\overline{t}^{2}-d\overline{x}%
^{2}] \label{eMD}%
\end{equation}
However, we note that the new radiation function $\overline{\phi}$ in Eq.
(\ref{F1}) can be regarded as a function of the old time and space coordinates
$\overline{\phi}(ct,x)=\operatorname{Re}\exp[i(k/\sigma)x-i(|k|/\sigma)ct)$
giving a new radiation field in the old coordinate system. \ We are interested
in time-dilating conformal changes in this sense of mapping normal modes into
new normal modes without any change in the coordinate system. \ Thus the
time-dilating conformal transformation can be regarded as mapping the normal
mode at frequency $c|k|$ into the normal mode at frequency $c|\sigma k|$
according to
\begin{equation}
\phi_{k}(ct,x)\rightarrow\phi_{\sigma k}(ct,x)=\phi_{k}(c\sigma t,\sigma
x)=\operatorname{Re}\exp[i\sigma kx-i|\sigma k|ct] \label{E1}%
\end{equation}
If the positive dilatation constant $\sigma$ is chosen larger than $1,$
$\sigma>1,$ then the energy density of the transformed normal mode
$\phi_{\sigma k}(ct,x)$ will be larger than the energy density of the original
normal mode $\phi_{k}(ct,x).$ \ 

We emphasize that the coordinate transformation from an inertial frame to a
Rindler frame is not conformal, and the wave equation (\ref{eWR1}) in a
Rindler frame is not the same as the wave equation (\ref{W1}) in an inertial frame.

It is clear that in an inertial frame, a space-dilating conformal
transformation coincides with a time-dilating conformal transformation; both
correspond to uniform dilations of all coordinates, and provide a
$\sigma_{ltU^{-1}}$-scale transformation. \ \ However, in a general coordinate
frame, a time-dilating conformal transformation is different from a
space-dilating conformal transformation. Thus for a Rindler frame in two
spacetime dimensions, a space-dilating conformal transformation involves the
transformation
\begin{equation}
\eta\rightarrow\overline{\eta}=\eta,\text{ \ \ }\xi\rightarrow\overline{\xi
}=\sigma\xi,\text{ \ \ }\varphi\rightarrow\overline{\varphi}(\overline{\eta
},\overline{\xi})=\varphi(\eta,\xi), \label{eCTs1}%
\end{equation}
giving the metric form%
\begin{equation}
ds^{2}=c^{2}dt^{2}-dx^{2}=\xi^{2}d\eta^{2}-d\xi^{2}=\sigma^{-2}(\overline{\xi
}^{2}d\overline{\eta}^{2}-d\overline{\xi}^{2}) \label{eCTs2}%
\end{equation}
\ Under the space-dilating conformal transformation (\ref{eCTs1}), we find the
wave equation transforms as%
\begin{align}
\frac{1}{\overline{\xi}^{2}}\frac{\partial^{2}\overline{\varphi}}%
{\partial\overline{\eta}^{2}}-\frac{\partial^{2}\overline{\varphi}}%
{\partial\overline{\xi}^{2}}-\frac{1}{\overline{\xi}}\frac{\partial
\overline{\varphi}}{\partial\overline{\xi}}  &  =\frac{1}{(\sigma\xi)^{2}%
}\frac{\partial^{2}\varphi}{\partial\eta^{2}}-\frac{\partial^{2}\varphi
}{\partial(\sigma\xi)^{2}}-\frac{1}{(\sigma\xi)}\frac{\partial\varphi
}{\partial(\sigma\xi)}\nonumber\\
&  =\frac{1}{\sigma^{2}}\left(  \frac{1}{\xi^{2}}\frac{\partial^{2}\varphi
}{\partial\eta^{2}}-\frac{\partial^{2}\varphi}{\partial\xi^{2}}-\frac{1}{\xi
}\frac{\partial\varphi}{\partial\xi}\right)  =0 \label{eWCs}%
\end{align}
giving the mapping of normal modes as%
\begin{align}
\overline{\varphi}(\overline{\eta},\overline{\xi})  &  =\varphi(\eta
,\xi)=\varphi(\overline{\eta},\overline{\xi}/\sigma)\nonumber\\
&  =\operatorname{Re}\exp[i\kappa\ln(\overline{\xi}/\sigma)-i|\kappa
|\overline{\eta}]\nonumber\\
&  =\operatorname{Re}\exp\left[  i\kappa\ln\overline{\xi}-i|\kappa
|\overline{\eta}-i\kappa\ln\sigma\right]  \label{R4}%
\end{align}
which is merely a change in the phase of the normal mode. \ 

On the other hand, a time-dilating conformal transformation corresponds to
\begin{equation}
\eta\rightarrow\overline{\eta}=\sigma\eta,\text{ \ \ }\xi\rightarrow
\overline{\xi}=\xi^{\sigma},\text{ \ \ }\varphi\rightarrow\overline{\varphi
}(\overline{\eta},\overline{\xi})=\varphi(\eta,\xi), \label{eCTs3}%
\end{equation}
giving the metric form%
\begin{equation}
ds^{2}=c^{2}dt^{2}-dx^{2}=\xi^{2}d\eta^{2}-d\xi^{2}=(\overline{\xi}%
^{(1/\sigma-1)}/\sigma)^{2}(\overline{\xi}^{2}d\overline{\eta}^{2}%
-d\overline{\xi}^{2}) \label{eCT2}%
\end{equation}
Under a time-dilating conformal transformation (\ref{eCTs3}), we find the wave
equation transforms as%
\begin{align}
\frac{1}{\overline{\xi}^{2}}\frac{\partial^{2}\overline{\varphi}}%
{\partial\overline{\eta}^{2}}-\frac{\partial^{2}\overline{\varphi}}%
{\partial\overline{\xi}^{2}}-\frac{1}{\overline{\xi}}\frac{\partial
\overline{\varphi}}{\partial\overline{\xi}}  &  =\frac{1}{(\xi^{\sigma})^{2}%
}\frac{\partial^{2}\varphi}{\partial\eta^{2}}-\frac{\partial^{2}\varphi
}{\partial(\xi^{\sigma})^{2}}-\frac{1}{(\xi^{\sigma})}\frac{\partial\varphi
}{\partial(\xi^{\sigma})}\nonumber\\
&  =\frac{1}{\sigma^{2}\xi^{2\sigma-2}}\left(  \frac{1}{\xi^{2}}\frac
{\partial^{2}\varphi}{\partial\eta^{2}}-\frac{\partial^{2}\varphi}{\partial
\xi^{2}}-\frac{1}{\xi}\frac{\partial\varphi}{\partial\xi}\right)  =0
\label{eWRT2}%
\end{align}
giving the mapping of normal modes $\varphi\rightarrow\overline{\varphi}$ as%
\begin{align}
\overline{\varphi}(\overline{\eta},\overline{\xi})  &  =\varphi(\eta
,\xi)=\varphi(\overline{\eta}/\sigma,\overline{\xi}^{1/\sigma})\nonumber\\
&  =\operatorname{Re}\exp[i\kappa\ln(\overline{\xi}^{1/\sigma})-i|\kappa
|\overline{\eta}/\sigma]\nonumber\\
&  =\operatorname{Re}\exp\left[  i\frac{\kappa}{\sigma}\ln\overline{\xi
}-i|\frac{\kappa}{\sigma}|\overline{\eta}\right]  \label{R5}%
\end{align}
which corresponds to a change in the "frequency" of the normal mode in the
Rindler frame. \ Again, we are interested in time-dilating conformal
transformations only in the sense of carrying normal mode solutions into
normal mode solutions. \ Thus the radiation field $\overline{\varphi}$ in Eq.
(\ref{R5}) can be regarded as a new radiation field in the original
coordinates $\overline{\varphi}(\eta,\xi)=\operatorname{Re}\exp[i(\kappa
/\sigma)\ln\xi-i\left\vert \kappa/\sigma\right\vert \eta]$ in the original
coordinate frame. \ We will treat a time-dilating conformal transformation in
a Rindler frame as the mapping of a Rindler normal mode $\varphi_{\kappa}%
(\eta,\xi)$ into a new normal mode $\varphi_{\sigma\kappa}(\eta,\xi)$ in the
same coordinate frame according to
\begin{equation}
\varphi_{\kappa}(\eta,\xi)\rightarrow\varphi_{\sigma\kappa}(\eta,\xi
)=\varphi_{\kappa}(\sigma\eta,\xi^{\sigma})=\operatorname{Re}\exp[i(\sigma
k)\xi-i|\sigma\kappa|\eta] \label{E5}%
\end{equation}
If the positive constant $\sigma$ exceeds $1,$ $\sigma>1,$ then the energy
density of the transformed normal mode $\varphi_{\sigma\kappa}(\eta,\xi)$ is
larger than the energy density of the initial normal mode $\varphi_{\kappa
}(\eta,\xi).$

\section{Zero-Point Radiation}

\subsection{Casimir Forces and Classical Zero-Point Radiation}

Experimental measurements of Casimir forces between conductors can be
described theoretically in terms of forces due to random classical radiation.
Experiment shows\cite{Spaarnay} that there are forces between uncharged
conductors which can be accounted for by assuming the presence of random
classical electromagnetic radiation even at zero temperature.\cite{CasimirF}
\ The spectrum of this zero-point radiation is uniquely determined by the
assumption that the spectrum of random classical radiation is Lorentz
invariant\cite{Marshall} in an inertial frame or, alternatively, is
$\sigma_{ltU^{-1}}$-scale invariant.\cite{conformal} The spectrum determined
in this fashion leads to theoretically-predicted forces in agreement with the
experimentally-measured separation-dependence of the Casimir forces, and the
one unknown scale factor in the classical zero-point radiation spectrum is
chosen to fit the magnitude of the experimental measurements. \ It is natural
to expect that thermal radiation at non-zero temperature $T>0$ fits smoothly
with the zero-point radiation at $T=0.$ \ We will make this smooth transition
under time-dilating conformal transformations the basis for obtaining the
thermal radiation spectrum within classical physics.

\subsection{Zero-Point Radiation in a General Spacetime}

It is a familiar idea that relativistic radiation contain no intrinsic
lengths, times, or energies, and so can be introduced into a general
spacetime. \ Classical zero-point radiation is random classical radiation
which has a scale-invariant spectrum. \ Since the radiation includes no
intrinsic lengths and the zero-point radiation spectrum also involves no
intrinsic lengths, the two-field correlation function for zero-point radiation
at spacetime points $P$ and $Q$ must involve only the distance along the
geodesic curve connecting $P$ and $Q,$ with a scale factor to be determined by
experiment. \ The scale factor is used to fit the experimentally-observed
Casimir forces between conductors which can be calculated within classical
physics based upon the classical electromagnetic zero-point radiation
spectrum. \ Within an inertial frame with its geodesic coordinates, the
spacetime separation squared between the spacetime points $P$ and $Q$ is given
by $s_{PQ}^{2}=c^{2}(t_{P}-t_{Q})^{2}-(\mathbf{r}_{P}-\mathbf{r}_{Q})^{2}$,
and the dimensionality of the fields will determine the zero-point correlation
function up to a multiplicative constant. \ Below we illustrate these ideas
for a scalar field in two spacetime dimensions. \ 

\subsection{Zero-Point Radiation in an Inertial Frame}

At this point, we obtain the forms taken by zero-point radiation in an
inertial frame and in a Rindler frame before we consider the transition to
thermal radiation. Zero-point radiation, and indeed thermal radiation at any
temperature, involves random radiation whose character does not change in
time. \ In any coordinate frame in which random radiation is a stationary
distribution, the radiation field can be expressed in terms of a sum over the
normal modes with random phases between the normal modes.\cite{Rice} \ In an
inertial frame in two spacetime dimensions, the correlation function for
time-stationary random radiation can be evaluated using plane wave normal
modes with random phases and requiring Lorentz invariance or scale invariance.
\ Thus a periodic normal mode $\phi_{k}(ct,x)$ of the scalar radiation field
in a box of length $l$ for $k=2n\pi/l$ (positive or negative) can be written
from Eqs. (\ref{M1}) and (\ref{eED1}) as
\begin{equation}
\phi_{k}(ct,x)=\left(  \frac{8\pi U(|k|)}{l}\right)  ^{1/2}\frac{1}{|k|}%
\cos[kx-|k|ct+\theta(k)] \label{eF1}%
\end{equation}
where $U(|k|)$ is the energy of the normal mode of angular frequency
$\omega=c|k|,$ and where $\theta(k)$ is a random phase which is distributed
uniformly on the interval $[0,2\pi)$ and is distributed independently for each
vector $k$. \ In the infinite-length limit, the random radiation field for
isotropic radiation can then be written as%
\begin{equation}
\phi(ct,x)=2%
%TCIMACRO{\dint _{-\infty}^{\infty}}%
%BeginExpansion
{\displaystyle\int_{-\infty}^{\infty}}
%EndExpansion
dk\,\frac{U^{1/2}(|k|)}{|k|}\cos[kx-|k|ct+\theta(k)] \label{R6}%
\end{equation}
When averaged over the random phases $\theta(k),$ we find\cite{volume}%
\begin{equation}
<\cos\theta(k)\cos\theta(k^{\prime})>=<\sin\theta(k)\sin\theta(k^{\prime
})>=\frac{1}{2}\delta(k-k^{\prime}) \label{eAV1}%
\end{equation}
and
\begin{equation}
<\sin\theta(k)\cos\theta(k^{\prime})>=0 \label{eAV2}%
\end{equation}
It follows that the two-field correlation function for isotropic random
radiation takes the form\cite{AJP2011}%
\begin{equation}
<\phi(ct,x)\phi(ct^{\prime},x^{\prime})>=2%
%TCIMACRO{\dint _{-\infty}^{\infty}}%
%BeginExpansion
{\displaystyle\int_{-\infty}^{\infty}}
%EndExpansion
dk\frac{U(|k|)}{k^{2}}\cos[k(x-x^{\prime})-|k|c(t-t^{\prime})] \label{ec3}%
\end{equation}
If we assume that the two-field correlation function is for zero-point
radiation $\phi_{0},$ and so is $\sigma_{ltU^{-1}}$-scale invariant under the
dilation transformation in two spacetime dimensions in an inertial frame where
the square of the field $\phi^{2}$ must transform as an invariant scalar
field, then we require
\begin{equation}
<\phi_{0}(c\sigma t,\sigma x)\phi_{0}(c\sigma t^{\prime},\sigma x^{\prime
})>=<\phi_{0}(ct,x),\phi_{0}(ct^{\prime},x^{\prime})> \label{R7}%
\end{equation}
In terms of the spectrum $U(|k|)$ appearing in Eq. (\ref{R6}), the
scale-invariance requirement is that%
\begin{align}
&  <\phi_{0}(c\sigma t,\sigma x)\phi_{0}(c\sigma t^{\prime},\sigma x^{\prime
})>\nonumber\\
&  =2%
%TCIMACRO{\dint }%
%BeginExpansion
{\displaystyle\int}
%EndExpansion
dk\frac{U(|k|)}{k^{2}}\cos[k\sigma(x-x^{\prime})-|k|c\sigma(t-t^{\prime
})]\nonumber\\
&  =2%
%TCIMACRO{\dint }%
%BeginExpansion
{\displaystyle\int}
%EndExpansion
\sigma dk\frac{\sigma U(|k|)}{(\sigma k)^{2}}\cos[k\sigma(x-x^{\prime
})-|k|c\sigma(t-t^{\prime})]\nonumber\\
&  =2%
%TCIMACRO{\dint }%
%BeginExpansion
{\displaystyle\int}
%EndExpansion
d\overline{k}\frac{(\sigma U(|\overline{k}/\sigma|)}{|\overline{k}|^{2}}%
\cos[\overline{k}(x-x^{\prime})-|\overline{k}|c(t-t^{\prime})] \label{ec4}%
\end{align}
should agree with $<\phi_{0}(ct,x)\phi_{0}(ct^{\prime},x^{\prime})>$ as in
Eq.(\ref{ec3}). This requirement means that
\begin{equation}
\sigma U(|k|/\sigma)=U(|k|) \label{ec6}%
\end{equation}
for all $k,$ which corresponds to $U(|k|)$ being linear in $|k|,$%
\begin{equation}
U(|k|)=\mathfrak{const}\times|k|. \label{ec7}%
\end{equation}
where $\mathfrak{const}$ is the multiplicative constant setting the scale of
classical electromagnetic zero-point radiation. \ This spectrum is also
Lorentz invariant\cite{AJP2011} so that zero-point radiation takes the same
spectral form in all inertial frames.

Having now found the spectrum (\ref{ec7}) for zero-point radiation in two
spacetime dimensions, we would like to evaluate the two-field correlation
function (\ref{ec3}) for zero-point radiation in closed form. \ However, the
correlation function is divergent at small values of $k.$ \ In order to avoid
this divergence, we will follow Davies and Fulling\cite{F-D} and will consider
the derivatives of the correlation function which take the form\cite{AJP2011}%
\begin{align}
&  <\phi_{0}(ct,x)\partial_{ct^{\prime}}\phi_{0}(ct^{\prime},x^{\prime})>=2%
%TCIMACRO{\dint }%
%BeginExpansion
{\displaystyle\int}
%EndExpansion
dk\left(  \frac{\mathfrak{const}}{|k|}\right)  (-|k|)\sin[\{k(x-x^{\prime
})-|k|c(t-t^{\prime})]\nonumber\\
&  =\mathfrak{const\times}\frac{-4c(t-t^{\prime})}{(x-x^{\prime})^{2}%
-c^{2}(t-t^{\prime})^{2}} \label{ed8}%
\end{align}
and%
\begin{align}
&  <\phi_{0}(ct,x)\partial_{x^{\prime}}\phi_{0}(ct^{\prime},x^{\prime})>=2%
%TCIMACRO{\dint }%
%BeginExpansion
{\displaystyle\int}
%EndExpansion
dk\left(  \frac{\mathfrak{const}}{|k|}\right)  (k)\sin[\{k(x-x^{\prime
})-|k|c(t-t^{\prime})]\nonumber\\
&  =\mathfrak{const\times}\frac{4(x-x^{\prime})}{(x-x^{\prime})^{2}%
-c^{2}(t-t^{\prime})^{2}} \label{ed9}%
\end{align}
Here the integrals involve a singular Fourier sine transform which is
evaluated using a temporary cut-off at high frequency.\cite{AJP2011}

\subsection{Zero-Point Radiation in \ a Rindler Frame}

Now we wish to consider zero-point radiation in a Rindler frame. \ Since the
relativistic scalar field is a scalar under coordinate transformations, taking
the same value at the same spatial point, $\varphi(\eta,\xi)=\phi
(ct,x)=\phi(\xi\sinh\eta,\xi\cosh\eta)$, we can obtain the two-field
correlation function for the scalar field simply by carrying out a coordinate
transformation
\begin{align}
&  <\varphi_{0}(\eta,\xi)\partial_{\eta^{\prime}}\varphi_{0}(\eta^{\prime}%
,\xi^{\prime})>\nonumber\\
&  =<\phi_{0}(ct,x)\partial_{ct^{\prime}}\phi_{0}(ct^{\prime},x^{\prime
})>\frac{\partial ct^{\prime}}{\partial\eta^{\prime}}+<\phi_{0}(ct,x)\partial
_{x^{\prime}}\phi_{0}(ct^{\prime},x^{\prime})>\frac{\partial x^{\prime}%
}{\partial\eta^{\prime}}\nonumber\\
&  =\mathfrak{const}\times\frac{-4c(t-t^{\prime})}{(x-x^{\prime})^{2}%
-c^{2}(t-t^{\prime})^{2}}\xi^{\prime}\cosh\eta^{\prime}+\mathfrak{const}%
\times\frac{4(x-x^{\prime})}{(x-x^{\prime})^{2}-c^{2}(t-t^{\prime})^{2}}%
\xi^{\prime}\sinh\eta^{\prime}\nonumber\\
&  =\mathfrak{const}\times\frac{4\xi^{\prime}(\xi\sinh\eta-\xi^{\prime}%
\sinh\eta^{\prime})\cosh\eta^{\prime}-4\xi^{\prime}(\xi\cosh\eta-\xi^{\prime
}\cosh\eta^{\prime})\sinh\eta^{\prime}}{[(\xi\sinh\eta-\xi^{\prime}\sinh
\eta^{\prime})^{2}-(\xi\cosh\eta-\xi^{\prime}\cosh\eta^{\prime})^{2}%
]}\nonumber\\
&  =\mathfrak{const}\times\frac{4\xi\xi^{\prime}\sinh(\eta-\eta^{\prime}%
)}{[2\xi\xi^{\prime}\cosh(\eta-\eta^{\prime})-\xi^{2}-\xi^{\prime2}]}
\label{ecR}%
\end{align}
and%

\begin{align}
&  <\varphi_{0}(\eta,\xi)\partial_{\xi^{\prime}}\varphi_{0}(\eta^{\prime}%
,\xi^{\prime})>\nonumber\\
&  =<\phi_{0}(ct,x)\partial_{ct^{\prime}}\phi_{0}(ct^{\prime},x^{\prime
})>\frac{\partial ct^{\prime}}{\partial\xi^{\prime}}+<\phi_{0}(ct,x)\partial
_{x^{\prime}}\phi_{0}(ct^{\prime},x^{\prime})>\frac{\partial x^{\prime}%
}{\partial\xi^{\prime}}\nonumber\\
&  =\mathfrak{const}\times\frac{-4c(t-t^{\prime})}{(x-x^{\prime})^{2}%
-c^{2}(t-t^{\prime})^{2}}\sinh\eta^{\prime}+\mathfrak{const}\times
\frac{4(x-x^{\prime})}{(x-x^{\prime})^{2}-c^{2}(t-t^{\prime})^{2}}\cosh
\eta^{\prime}\nonumber\\
&  =\mathfrak{const}\times\frac{4(\xi\sinh\eta-\xi^{\prime}\sinh\eta^{\prime
})\sinh\eta^{\prime}-4(\xi\cosh\eta-\xi^{\prime}\cosh\eta^{\prime})\cosh
\eta^{\prime}}{[(\xi\sinh\eta-\xi^{\prime}\sinh\eta^{\prime})^{2}-(\xi
\cosh\eta-\xi^{\prime}\cosh\eta^{\prime})^{2}}\nonumber\\
&  =\mathfrak{const}\times\frac{4[\xi^{\prime}-\xi\cosh(\eta-\eta^{\prime}%
)]}{[2\xi\xi^{\prime}\cosh(\eta-\eta^{\prime})-\xi^{2}-\xi^{\prime2}]}
\label{ecR2}%
\end{align}
We notice immediately that this correlation function is time stationary since
it depends only upon the time difference $\eta-\eta^{\prime}$ and not on the
individual times $\eta$ and $\eta^{\prime}.$ \ Indeed we expect that
zero-point radiation is the only spectrum of random radiation which is time
stationary in both all inertial frames and in all Rindler frames.

\section{Thermal Radiation}

\subsection{Thermal Radiation at $T>0$ is Coordinate-Frame Dependent}

Although the correlation function for classical zero-point radiation can be
specified in a coordinate-free manner as involving the separation along a
geodesic curve between the spacetime points where the fields are evaluated,
thermal radiation at non-zero temperature $T>0$ requires a specific coordinate
frame. \ Thus although zero-point radiation is Lorentz-invariant in an
inertial frame, thermal radiation at $T>0$ involves a finite (not divergent)
density of additional energy above the divergent zero-point spectrum, and so
is associated with a unique coordinate frame, namely the frame in which the
random radiation is time-stationary and in which radiation is isotropic at
every spatial point.

\subsection{Thermal Radiation and Inertial-Frame Dilatations of the Conformal
Group}

In an inertial frame, the pattern of thermal radiation depends upon the one
parameter of temperature $T$. \ Since the conformal transformations are
symmetry transformations of electromagnetism, we might expect some interesting
transformations when conformal transformations are applied to the thermal
radiation spectrum. \ In an inertial frame, the dilatations of the conformal
group are uniform scale transformations of length, time, and energy while the
proper conformal transformations can be regarded as local spacetime-dependent
scale transformations.\cite{Kastrup} \ Under dilatations of the conformal
group in an inertial frame, thermal radiation at temperature $T>0$ is carried
into thermal radiation at a new temperature $T/\sigma$ which is a positive
multiple of the old temperature and which can approach ever close to zero
temperature as the positive constant $\sigma\rightarrow\infty$. \ However,
once zero temperature is reached in an inertial frame, the process can not be
reversed by use of the dilatations of the conformal group; one can not map
zero-point radiation into non-zero thermal radiation by conformal-group
dilatations in an inertial frame.

\ This scale invariance of zero-point radiation in an inertial frame was
discussed in Section III C, but here we wish to reiterate the invariance in
connection with the closed form expressions for the correlation functions
given in (\ref{ed8}) and (\ref{ed9}). \ Thus the invariance follows from
$\partial t/\partial(\sigma t)=1/\sigma$ and $\partial x/\partial(\sigma
x)=1/\sigma$, where%

\begin{align}
&  <\phi_{0}(c\sigma t,\sigma x)\partial_{c\sigma t^{\prime}}\phi_{0}(c\sigma
t^{\prime},\sigma x^{\prime})>=\mathfrak{const}\frac{-4c(\sigma t-\sigma
t^{\prime})}{(\sigma x-\sigma x^{\prime})^{2}-c^{2}(\sigma t-\sigma t^{\prime
})^{2}}\nonumber\\
&  =\frac{1}{\sigma}\mathfrak{const}\frac{-4c(t-t^{\prime})}{(x-x^{\prime
})^{2}-c^{2}(t-t^{\prime})^{2}}=\frac{1}{\sigma}<\phi_{0}(ct,x)\partial
_{ct^{\prime}}\phi_{0}(ct^{\prime},x^{\prime})> \label{R8}%
\end{align}%
\begin{align}
&  <\phi_{0}(c\sigma t,\sigma x)\partial_{x^{\prime}}\phi_{0}(c\sigma
t^{\prime},\sigma x^{\prime})>=\mathfrak{const}\frac{4(\sigma x-\sigma
x^{\prime})}{(\sigma x-\sigma x^{\prime})^{2}-c^{2}(\sigma t-\sigma t^{\prime
})^{2}}\nonumber\\
&  =\frac{1}{\sigma}\mathfrak{const}\frac{4(x-x^{\prime})}{(x-x^{\prime}%
)^{2}-c^{2}(t-t^{\prime})^{2}}=\frac{1}{\sigma}<\phi_{0}(ct,x)\partial
_{cx^{\prime}}\phi_{0}(ct^{\prime},x^{\prime})> \label{R9}%
\end{align}
\ We see again that zero-point radiation is completely invariant under the
dilatations of the conformal group in an inertial frame.

\subsection{Contrasting Behavior for Thermal Radiation in Inertial and
Non-Inertial Frames}

An inertial frame is a very special system, not only as far as geodesic
coordinates are concerned but also as far as thermal radiation is concerned.
\ An inertial frame has no intrinsic lengths or times which are associated
with the spatial points of the frame. \ Accordingly, in equilibrium, the
temperature, pressure, energy density, and field correlations are everywhere
the same throughout the entire spacetime. The situation is completely
different in a non-inertial frame, for example in a Rindler frame. \ In a
non-inertial frame, the temperature at equilibrium will in general be a local
function of the spatial coordinates, just as the local proper time at a fixed
coordinate is a function of the spatial coordinates and the time parameter.
\ Thus in a Rindler frame, the equilibrium temperature $T$ is not constant but
rather varies inversely as the distance $\xi$ to the event horizon; the
product $T\xi$ is a constant, and from Eq. (\ref{eMR}) in a Rindler frame, the
proper time interval $d\tau$ at a fixed spatial coordinate $\xi$ is given by
$cd\tau=ds=\xi d\eta.$ \ The contrast in behavior between inertial and
non-inertial frames suggests the possibility that time-dilating conformal
transformations may connect zero-point radiation continuously with thermal
radiation at non-zero-temperature in a non-inertial frame. \ Associated with
the coordinates in a non-inertial frame will be characteristic times which are
related to variations in the zero-point radiation correlation function.
\ Under a time-dilating conformal transformation, the spectrum of zero-point
radiation will be carried into a new stationary distribution of random
radiation which is different from zero-point radiation. \ If the zero-point
radiation spectrum is connected continuously with non-zero temperature
radiation, then we expect a time-dilating conformal transformation to carry
the zero-point spectrum into the thermal spectrum at non-zero-temperature in a
non-inertial frame.

\subsection{Space-Dilating Conformal Transformations in a Rindler Frame}

The dilatations associated with conformal symmetry can be carried out easily
in a Rindler frame in two spacetime dimensions. \ If we first consider the
space-dilating conformal transformation of Eq. (\ref{eCTs1}), then we find
\ that zero-point radiation in a Rindler frame is invariant. \ Thus from
$\partial\xi^{\prime}/\partial(\sigma\xi^{\prime})=1/\sigma,$ we have
\begin{align}
&  <\varphi_{0}(\eta,\sigma\xi)\partial_{\eta^{\prime}}\varphi_{0}%
(\eta^{\prime},\sigma\xi^{\prime})>=\mathfrak{const}\frac{4(\sigma\xi
)(\sigma\xi^{\prime})\sinh(\eta-\eta^{\prime})}{[2(\sigma\xi)(\sigma
\xi^{\prime})\cosh(\eta-\eta^{\prime})-(\sigma\xi)^{2}-(\sigma\xi^{\prime
})^{2}]}\nonumber\\
&  =\mathfrak{const}\frac{4\xi\xi^{\prime}\sinh(\eta-\eta^{\prime})}{[2\xi
\xi^{\prime}\cosh(\eta-\eta^{\prime})-\xi^{2}-\xi^{\prime2}]}=<\varphi
_{0}(\eta,\xi)\partial_{\eta^{\prime}}\varphi_{0}(\eta^{\prime},\xi^{\prime})>
\label{S1}%
\end{align}
while%
\begin{align}
&  <\varphi_{0}(\eta,\sigma\xi)\partial_{\sigma\xi^{\prime}}\varphi_{0}%
(\eta^{\prime},\sigma\xi^{\prime})>=\mathfrak{const}\frac{4[(\sigma\xi
^{\prime})-(\sigma\xi)\cosh(\eta-\eta^{\prime})]}{[2(\sigma\xi)(\sigma
\xi^{\prime})\cosh(\eta-\eta^{\prime})-(\sigma\xi)^{2}-(\sigma\xi^{\prime
})^{2}]}\nonumber\\
&  =\mathfrak{const}\frac{4[\xi^{\prime}-\xi\cosh(\eta-\eta^{\prime})]}%
{[2\xi\xi^{\prime}\cosh(\eta-\eta^{\prime})-\xi^{2}-\xi^{\prime2}]}=\frac
{1}{\sigma}<\varphi_{0}(\eta,\xi)\partial_{\xi^{\prime}}\varphi_{0}%
(\eta^{\prime},\xi^{\prime})> \label{S2}%
\end{align}
From the metric in Eq. (\ref{eMR}), we see that the Rindler spatial coordinate
$\xi$ is a geodesic coordinate and so the space-dilating conformal
transformation leaves the zero-point correlation function unchanged. \ This
invariance is consistent with the result in Eq. (\ref{R4}) which showed that
the space-dilating conformal transformation in a Rindler frame merely changed
the phase of a radiation normal mode. \ Since zero-point radiation has random
phases between the modes, this change in phase for each mode does not alter
the correlation function for zero-point radiation in the Rindler frame.

\subsection{Time-Dilating Conformal Transformations in a Rindler Frame}

Next we consider a time-dilating conformal transformation of the zero-point
radiation in the Rindler frame. \ The time-dilating conformal transformation
is given in Eq. (\ref{eCTs3}). \ In this case, we find
\begin{equation}
<\varphi_{0}(\sigma\eta,\xi^{\sigma})\partial_{\sigma\eta^{\prime}}\varphi
_{0}(\sigma\eta^{\prime},\xi^{\prime\sigma})>=\mathfrak{const}\frac
{4\xi^{\sigma}\xi^{\prime\sigma}\sinh(\sigma\eta-\sigma\eta^{\prime})}%
{[2\xi^{\sigma}\xi^{\prime\sigma}\cosh(\sigma\eta-\sigma\eta^{\prime}%
)-\xi^{\sigma2}-\xi^{\prime\sigma2}]} \label{S3}%
\end{equation}%
\begin{equation}
<\varphi_{0}(\sigma\eta,\xi^{\sigma})\partial_{\xi^{\prime^{s}}}\varphi
_{0}(\sigma\eta^{\prime},\xi^{\prime\sigma})>=\mathfrak{const}\frac
{4[\xi^{\prime\sigma}-\xi^{\sigma}\cosh(\sigma\eta-\sigma\eta^{\prime}%
)]}{[2\xi^{\sigma}\xi^{\prime\sigma}\cosh(\sigma\eta-\sigma\eta^{\prime}%
)-\xi^{\sigma2}-\xi^{\prime\sigma2}]} \label{S4}%
\end{equation}
These correlation functions (\ref{S3}) and (\ref{S4}) found from the
time-dilating conformal transformation are not invariant but rather involve
entirely new functional forms, different from the correlations for zero-point
radiation in Eqs. (\ref{ecR}) and (\ref{ecR2}). \ 

In order to simplify the situation, we consider the correlation functions in
Eqs. (\ref{S3}) and (\ref{S4}) at different times $\eta$ and $\eta^{\prime}$
but at a single spatial point $\xi=\xi^{\prime}.$ \ The correlation function
(\ref{S4}) involving the spatial derivative becomes%
\begin{equation}
<\varphi_{0}(\sigma\eta,\xi^{\sigma})\partial_{\xi^{\prime^{s}}}\varphi
_{0}(\sigma\eta^{\prime},\xi^{\prime\sigma})>_{\xi=\xi^{\prime}}%
=-\mathfrak{const}\frac{2}{\xi^{\sigma}} \label{S5}%
\end{equation}
and is independent of the time parameter $\eta$ or $\eta^{\prime}.$ \ The
correlation function (\ref{S3}) involving the time derivative becomes%
\begin{align}
&  <\varphi_{0}(\sigma\eta,\xi^{\sigma})\partial_{\sigma\eta^{\prime}}%
\varphi_{0}(\sigma\eta^{\prime},\xi^{\sigma})>=\mathfrak{const}\frac
{4\xi^{\sigma}\xi^{\sigma}\sinh(\sigma\eta-\sigma\eta^{\prime})}{[2\xi
^{\sigma}\xi^{\sigma}\cosh(\sigma\eta-\sigma\eta^{\prime})-\xi^{\sigma2}%
-\xi^{\sigma2}]}\nonumber\\
&  =\mathfrak{const}\frac{4\sinh[\sigma(\eta-\eta^{\prime})]}{2\cosh
[\sigma(\eta-\eta^{\prime})]-2}=\mathfrak{const}\frac{8\sinh[\sigma(\eta
-\eta^{\prime})/2]\cosh[\sigma(\eta-\eta^{\prime})/2]}{2\{2\sinh^{2}%
[\sigma(\eta-\eta^{\prime})/2]\}}\nonumber\\
&  =\mathfrak{const}\times2\coth[\sigma(\eta-\eta^{\prime})/2] \label{S6}%
\end{align}
which is independent of the spatial point $\xi.$ \ Now as discussed in Section
II E, we are interpreting the time-dilating conformal transformation as a
transformation of the radiation field in the same coordinate frame. \ Thus in
Eq. (\ref{S6}), this correlation function obtained from zero-point radiation
by a time-dilating conformal transformation should involve the correlation
function $<\varphi_{\alpha}(\eta,\xi)\varphi_{\alpha}(\eta^{\prime},\xi)>$ at
times $\eta$ and $\eta^{\prime}$ for thermal radiation at some finite non-zero
temperature with temperature parameter $\alpha,$ corresponding to
\begin{equation}
<\varphi_{\alpha}(\eta,\xi)\varphi_{\alpha}(\eta^{\prime},\xi)>=<\varphi
_{0}(\sigma\eta,\xi^{\sigma})\varphi_{0}(\sigma\eta^{\prime},\xi^{\sigma})>
\label{E9}%
\end{equation}
in analogy with the time-dilating conformal transformation in Eq. (\ref{E5}).
\ If we introduce the derivative with respect to $\eta^{\prime}$ so as to make
direct contact with Eq. (\ref{S6}), then we have%
\begin{align}
\frac{\partial}{\partial(\xi\eta^{\prime})}  &  <\varphi_{\alpha}(\eta
,\xi)\varphi_{\alpha}(\eta^{\prime},\xi)>=\frac{\sigma}{\xi}\frac{\partial
}{\partial(\sigma\eta^{\prime})}<\varphi_{0}(\sigma\eta,\xi^{\sigma}%
)\varphi_{0}(\sigma\eta^{\prime},\xi^{\sigma})>\nonumber\\
&  =\frac{\sigma}{\xi}<\varphi_{0}(\sigma\eta,\xi^{\sigma})\partial
_{\sigma\eta^{\prime}}\varphi_{0}(\sigma\eta^{\prime},\xi^{\sigma
})>\nonumber\\
&  =\mathfrak{const}\frac{2\sigma}{\xi}\coth[\sigma(\eta-\eta^{\prime
})/2]=\mathfrak{const}\frac{2\sigma}{\xi}\coth\left(  \frac{\sigma}{\xi}%
\frac{(\xi\eta-\xi\eta^{\prime})}{2}\right)  \label{D1}%
\end{align}
If we now introduce the proper time $\tau=\xi\eta/c$ at the fixed spatial
point $\xi$ in the Rindler frame, then the correlation function (\ref{D1})
becomes%
\begin{align}
\frac{\partial}{\partial(\xi\eta^{\prime})}  &  <\varphi_{\alpha}(\eta
,\xi)\varphi_{\alpha}(\eta^{\prime},\xi)>=\frac{\partial}{\partial
c\tau^{\prime}}<\varphi_{\alpha}(\eta,\xi)\varphi_{\alpha}(\eta^{\prime}%
,\xi)>\nonumber\\
&  =\mathfrak{const}\frac{2\sigma}{\xi}\coth\left(  \frac{\sigma}{\xi}%
\frac{(c\tau-c\tau^{\prime})}{2}\right)  \label{S7}%
\end{align}
where $\tau-\tau^{\prime}$ is the proper time difference between the spacetime
points and $\sigma/\xi$ corresponds to some parameter giving the temperature
of the radiation at the location $\xi.$ \ 

\subsection{Thermal Radiation in an Inertial Frame Obtained as a Limit from a
Rindler Frame}

We obtain the correlation function for thermal radiation in an inertial frame
from the thermal radiation correlation function (\ref{S7}) by moving the
spatial point $\xi$ ever further from the event horizon, $\xi\rightarrow
\infty,$ but keeping the temperature parameter $\sigma/\xi$ constant by
continuously increasing $\sigma$. \ In the limit, we obtain the time
correlation function for thermal radiation in vanishing proper acceleration
$c^{2}/\xi\rightarrow0,$ which corresponds to thermal radiation in an inertial
frame\cite{AJP2011}%
\begin{align}
\frac{\partial}{\partial ct^{\prime}}  &  <\phi_{\alpha}(ct,x)\phi_{\alpha
}(ct^{\prime},x)>\nonumber\\
&  =\mathfrak{const}\frac{2\sigma}{\xi}\coth\left(  \frac{\sigma}{\xi}%
\frac{(ct-ct^{\prime})}{2}\right)  \label{H1}%
\end{align}

The actual radiation spectrum is obtained from the correlation function Eq.
(\ref{H1}) by taking the inverse Fourier time transform of the correlation
function expression (\ref{ec3}),
\begin{align}
&  <\phi_{\alpha}(ct,x)\partial_{ct^{\prime}}\phi_{\alpha}(ct^{\prime},x)>=2%
%TCIMACRO{\dint _{-\infty}^{\infty}}%
%BeginExpansion
{\displaystyle\int_{-\infty}^{\infty}}
%EndExpansion
dk\frac{U_{T}(|k|)}{k^{2}}(-|k|)\sin[k(x-x^{\prime})-|k|c(t-t^{\prime
})]_{x=x^{\prime}}\nonumber\\
&  =2%
%TCIMACRO{\dint _{-\infty}^{\infty}}%
%BeginExpansion
{\displaystyle\int_{-\infty}^{\infty}}
%EndExpansion
dk\frac{U_{T}(|k|)}{k^{2}}(|k|)\sin[|k|c(t-t^{\prime})]=\mathfrak{const}%
\frac{2\sigma}{\xi}\coth\left(  \frac{\sigma}{\xi}\frac{(ct-ct^{\prime})}%
{2}\right)  \label{H2}%
\end{align}
giving the energy $U_{T}(|k|)$ per normal mode of frequency $\omega=c|k|$ at
some temperature $T$ as\cite{AJP2011}%
\begin{equation}
U_{T}(|k|)=\mathfrak{const}\times|k|\coth\left(  \frac{|k|\pi}{(\sigma/\xi
)}\right)  \label{S9}%
\end{equation}
The temperature $T$ can be obtained by going to the high-temperature or
low-frequency limit where the Rayleigh-Jeans spectrum becomes valid, giving%
\begin{equation}
U_{T}(|k|)\rightarrow\mathfrak{const}\times|k|\left(  \frac{|k|\pi}%
{(\sigma/\xi)}\right)  ^{-1}=\mathfrak{const}\frac{\sigma}{\pi\xi}=k_{B}T
\label{T1}%
\end{equation}
Thus the spectrum of thermal radiation is given by%
\begin{equation}
U_{T}(|k|)=\mathfrak{const}\times|k|\coth\left(  \frac{\mathfrak{const}%
\times|k|}{k_{B}T}\right)  \label{T2}%
\end{equation}
Traditionally, the spectrum of blackbody thermal radiation is given in terms
of Planck's constant $\hbar$ rather than in terms of the scale factor
$\mathfrak{const}$ for classical electromagnetic zero-point radiation given in
Eq. (\ref{ec7}). \ Since the connection involves%
\begin{equation}
\mathfrak{const}=\frac{1}{2}\hbar c \label{T3}%
\end{equation}
our equation (\ref{T2}) corresponds to the familiar Planck spectrum with
zero-point radiation,
\begin{equation}
U(\omega,T)=\frac{1}{2}\mathfrak{\hbar}\omega\coth\left(  \frac
{\mathfrak{\hbar}\omega}{2k_{B}T}\right)  =\frac{1}{2}\hbar\omega+\frac
{\hbar\omega}{\exp[\hbar\omega/(k_{B}T)]-1} \label{T4}%
\end{equation}
We emphasize that in the limit $T\rightarrow0,$ the spectrum $U(\omega,T)$ of
classical random radiation in Eq. (\ref{T4}) does not vanish, but rather goes
over to the Lorentz-invariant spectrum of classical zero-point radiation.
\ The Planck spectrum is connected continuously with zero-point radiation.
\ Within classical physics, this continuous evolution of the spectrum of
random radiation at non-zero temperature into the zero-point radiation
spectrum is consistent with the results for Casimir forces where the forces at
non-zero temperature are connected smoothly with the experimentally-observed
forces at low temperature.

\section{Discussion}

Thermal radiation is the random radiation which appears inside an isolated
enclosure, and this radiation, like all random radiation, causes Casimir
forces between conducting plates and dielectric surfaces, associated with the
discrete nature of radiation normal modes in restricted volumes. \ At high
temperatures, the random thermal radiation produces Casimir forces which
follow from the Rayleigh-Jeans spectrum and depend only on the temperature $T$
and the geometrical dimensions.\cite{PR1975} \ However at low temperatures as
$T\rightarrow0,$ the Casimir forces do not vanish, indicating the presence of
random radiation even at zero temperature. \ This radiation corresponds to
classical zero-point radiation. \ We expect a smooth evolution of the thermal
radiation spectrum with temperature as the temperature is lowered, fitting
with a smooth evolution of the Casimir forces at high temperature over to the
Casimir forces at zero temperature. \ Under a time-dilating conformal
transformation, thermal radiation is carried into itself, so that at each
spatial point $T\rightarrow T^{\prime}=T/\sigma$ where $\sigma$ is a positive
constant. \ Thermal radiation is also a situation of greatest randomness and
least information. \ At zero temperature, the least information means that the
random radiation gives no more information than is contained in the structure
of spacetime itself. \ Thus at zero temperature, we expect that the two-field
zero-point correlation function at points $P$ and $Q$\ involves only the
distance along a geodesic between the spacetime points. \ In a noninertial
static coordinate frame, the time parameter is separated from the spatial
coordinates so that a time-dilating conformal transformation can carry
zero-point radiation into thermal radiation at non-zero temperature. This
general situation is quite different from that in an inertial frame where a
time-dilating conformal transformation is the same as a uniform coordinate
dilation, and the two-field zero-point correlation function is invariant under
a conformal transformation. \ Thus in an inertial frame, information about
zero-point radiation does not give us any information about the thermal
radiation spectrum at finite non-zero temperature. \ In this article, we have
considered classical random scalar radiation in a Rindler frame in two
spacetime dimensions. \ We find that a time-dilating conformal transformation
carries the zero-point radiation spectrum into the thermal radiation spectrum
at a non-zero temperature in a Rindler frame. \ The thermal radiation spectrum
in the Rindler frame can be carried back to an inertial frame, giving the
Planck spectrum with zero-point radiation. \ Thus Planck's spectrum of
blackbody radiation follows from zero-point radiation and the structure of
relativistic spacetime in classical physics.

\section{Acknowledgement}

I would like to thank Professor Parameswaran Nair for helpful conversations
regarding conformal symmetry.


\begin{thebibliography}{99}                                                                                               %


\bibitem {PR2010}T. H. Boyer, "Derivation of the Planck spectrum for
relativistic classical scalar radiation from thermal equilibrium in an
accelerating frame," Phys. Rev. D \textbf{81}, 105024 (2010).

\bibitem {AJP2011}T. H. Boyer, "Classical physics of thermal scalar radiation
in two spacetime dimensions," Am. J. Phys. \textbf{79}, 644-656 (2011).

\bibitem {Eisberg}See for example, R. Eisberg and R. Resnick, \textit{Quantum
Physics of Atoms, Molecules, Solids, Nuclei, and Particles 2nd ed.} (Wiley,
New York 1985).

\bibitem {FOP2010}T. H. Boyer, "Blackbody radiation and the scaling symmetry
of relativistic classical electron theory with classical electromagnetic
zero-point radiation," Found. Phys. \textbf{40}, 1102-1116 (2010).

\bibitem {VanV}J. H. van Vleck, "The absorption of radiation by multiply
periodic orbits, and its relation to the correspondence principle and the
Rayleigh-Jeans law: Part II. Calculation of absorption by multiply periodic
orbits," Phys. Rev. \textbf{24}, 347-365 (1924). \ See also, T. H. Boyer,
"Equilibrium of random classical electromagnetic radiation in the presence of
a nonrelativistic nonlinear electric dipole oscillator," Phys. Rev. D
\textbf{13}, 2832-2845 (1976); "Statistical equilibrium of nonrelativistic
multiply periodic classical systems and random classical electromagnetic
radiation," Phys. Rev. A \textbf{18}, 1228-1237 (1978).

\bibitem {BC1909}E. Cunningham, "The principle of relativity in
electrodynamics and an extension thereof ," Proc. London Math. Soc.
\textbf{8}, 77-98 \ (1910); H. Bateman, "the transformation of the
electrodynamical equations," Proc. London Math. Soc. \textbf{8}, 223-264 \ (1910).

\bibitem {Kastrup}H. A. Kastrup, "Zur physikalischen Deuting und
darstellungstheoretischen Analyse der konformen Transformationen der Raum und
Zeit," Ann. Phys. (Leipzig) \textbf{9}, 388-428 (1962). \ 

\bibitem {SCALING}See the discussion by T. H. Boyer, "Scaling symmetry and
thermodynamic equilibrium for classical electromagnetic radiation," Found.
Phys. \textbf{19}, 1371-1383 (1989), and also in ref. 2.

\bibitem {Rindler}See for example, W. Rindler, \textit{Essential Relativity:
Special, General, and Cosmological 2nd ed.} (Springer-Verlag, New York 1977),
pp. 49-51. \ See also, See for example, B. F. Schutz, \textit{A First Course
in General Relativity} (Cambridge U. Press 1986), p. 150.

\bibitem {Rohrlich}T. Fulton, F. Rohrlich, and L. Witten, "Conformal
invariance in physics," Rev. Mod. Phys. \textbf{34}, 442-457 (1962).

\bibitem {Kastrup2}It has been pointed out by Kastrup in ref. 7 that a
conformal transformation in an inertial frame can be regarded as a smooth
$\sigma_{ltU^{-1}}$-scale transformation with a varying scale change from
point to point. \ This idea is consistent with the observation that
relativistic radiation determines the metric of a spacetime only up to a scale
which can vary continuously from point to point.

\bibitem {Spaarnay}M. J. Sparnaay, "Measurement of the attractive forces
between flat plates," Physica (Amsterdam) \textbf{24}, 751-764 (1958); S. K.
Lamoreaux, "Demonstration of the Casimir force in the 0.6 to 6 $\mu$m range,"
Phys. Rev. Lett. \textbf{78}, 5-8 (1997): \textbf{81}, 5475-5476 (1998); U.
Mohideen, "Precision measurement of the Casimir force from 0.1 to 0.9 $\mu$m,"
\textit{ibid.} \textbf{81}, 4549-4552 (1998); H. B. Chan, V. A. Aksyuk, R. N.
Kleinman, D. J. Bishop, and F. Capasso, "Quantum mechanical actuation of
microelectromechanical systems by the Casimir force," Science \textbf{291},
1941-1944 (2001): G. Bressi, G. Caarugno, R. Onofrio, and G. Ruoso,
"Measurement of the Casimir force between parallel metallic surfaces," Phys.
Rev. Lett. \textbf{88}, 041804(4) (2002).

\bibitem {CasimirF}The original calculation was made in terms of the
zero-point energy of quantum field theory by H. B. G. Casimir, "On the
attraction between two perfectly conducting plates," Proc. Ned. Akad.
Wetenschap. \textbf{51}, 793-795 (1948). \ However, the same results appear in
classical electrodynamics which includes classical electromagnetic zero-point
radiation. \ See for example, T. H. Boyer, "Random electrodynamics: The theory
of classical electrodynamics with classical electromagnetic zero-point
radiation," Phys. Rev. \textbf{11}, 790-808 (1976).

\bibitem {Marshall}T. W. Marshall, "Statistical Electrodynamics," Proc. Camb.
Phil. Soc. \textbf{61}, 537-546 (1965); T. H. Boyer, "Derivation of the
Blackbody Radiation Spectrum without Quantum Assumptions," Phys. Rev.
\textbf{182}, 1374-11383 (1969).

\bibitem {conformal}T. H. Boyer, "Conformal Symmetry of Classical
Electromagnetic Zero-Point Radiation," Found. Phys. \textbf{19}, 349-365 (1989).

\bibitem {Rice}S. O. Rice, in Selected Papers on Noise and Stochastic
Processes, edited by N. Wax (Dover, New York 1954), p. 133.

\bibitem {volume}Here we are assuming the infinite-length limit.

\bibitem {F-D}S. A. Fulling and P. C. W. Davies, "Radiation from a moving
mirror in two dimensional space-time: conformal anomaly," Proc. R. Soc. Lond.
A. \textbf{348}, 393-414 (1976), p. 407.

\bibitem {PR1975}T. H. Boyer, "Temperature dependence of Van der Waals forces
in classical electrodynamics with classical electromagnetic zero-point
radiation," Phys. Rev. A \textbf{11}, 1650-1663 (1975).
\end{thebibliography}
\end{document}